\documentclass[floatfix,twocolumn,amsmath]{revtex4}
\pdfoutput=1
\usepackage{graphicx}

\usepackage{color}
\usepackage{subfigure}

\begin{document}

\title{Proteins with greater influence on network dynamics\\evolve more slowly but are not more essential}


\author{Ryan N. Gutenkunst}
\email{ryang@lanl.gov}
\affiliation{Theoretical Biology and Biophysics, Los Alamos National Laboratory}
\affiliation{Center for Nonlinear Studies, Los Alamos National Laboratory}

\date{\today}


\begin{abstract}
A fundamental question for evolutionary biology is why rates of evolution vary dramatically between proteins.
Perhaps surprisingly, it is controversial how much a protein's functional importance affects its rate of evolution.
In most studies, functional importance has been measured on the coarse scale of protein knock-outs, while evolutionary rate has been measured on the fine scale of amino acid substitutions.
Here we introduce dynamical influence, a finer measure of protein functional importance.
To measure dynamical influence, we first use detailed biochemical models of particular reaction networks to measure the influence of each reaction rate constant on network dynamics.
We then define the dynamical influence of a protein to be the average influence of the rate constants for all reactions it is involved in.

Using models of a dozen biochemical systems and sequence data from vertebrates, we show that dynamical influence and evolutionary rate are negatively correlated; proteins with greater dynamical influence evolve more slowly.
We also show that proteins with greater dynamical influence are not more likely to be essential.
This suggests that there are many cellular reactions whose presence is essential for life, but whose quantitative rate is relatively unimportant to fitness.
We also provide evidence that the effect of dynamical influence on evolutionary rate is independent of protein expression level, expression specificity, gene compactness, and reaction degree.
Dynamical influence offers a finer view of functional importance, and our results suggest that focusing on essentiality may have previously led to an underestimation of the role function plays in protein evolution.
\end{abstract}

\maketitle




\section*{Introduction}

Every protein evolves at a characteristic rate~\cite{bib:Zuckerkandl1965}, and these rates vary dramatically from protein to protein.
Understanding this variation is a fundamental challenge for evolutionary biology.
A predictive theory for evolutionary rates might also guide drug design toward targets that evolve slowly, lessening the problem of evolved drug resistance in pathogens.

The recent revolution in genomics has enabled many factors to be examined for correlation with protein evolutionary rate; among them are expression level~\cite{bib:Drummond2005}, gene compactness~\cite{bib:Liao2006}, knock-out essentiality~\cite{bib:Wang2009}, and interaction degree~\cite{bib:Fraser2002}.
Perhaps surprisingly, it is controversial how much a protein's functional importance impacts its rate of evolution~\cite{bib:Park2009}.
Theoretical considerations suggest that functional importance should play a substantial role~\cite{bib:Wilson1977}.
Studies in bacteria~\cite{bib:Jordan2002}, yeast~\cite{bib:Wang2009}, and mammals~\cite{bib:Liao2006} have shown that the essentiality of a gene, i.e. the fitness effect of gene deletion, has only a weak (but statistically significant) correlation with its evolutionary rate.

Protein evolutionary rate is most commonly by the ratio dN/dS of the rate of nonsynonymous (amino acid changing, dN) to synonymous (amino acid preserving, dS) genetic mutation.
Protein importance is most commonly assessed by the fitness effect of knocking that protein out.
Individual substitutions may, however, produce only small changes in a protein's structure and, consequently, small changes in its functional efficacy.
It is thus not clear that the fitness effect of a protein knock-out will correlate well with the fitness effect of small changes in protein efficacy, and this may explain the weak correlation seen between protein importance measured by knock-out and evolutionary rate.

Here we take advantage of another recent revolution in biology, the rise of mathematical modeling, to take an alternative and much finer view of a protein's functional importance.
We assess how small changes in the efficacy of a protein's reactions impact the dynamics of the network in which the protein is embedded.
To do so, we use computational models of specific molecular networks. 
Each of these models condenses the results of many biochemical studies, and we use them to probe the fine scale influence of variation in reaction rate constants on the network dynamics~\cite{bib:Gutenkunst2007}.
We construct a measure of a protein's influence on network dynamics from the influence of each constant for each reaction in which it is involved, and we correlate this with evolutionary rate.
The models we consider were all constructed in mice, rats, or humans, but we expect conservation among vertebrates of network structure~\cite{bib:Fokkens2009} and dynamics, allowing us to assess evolutionary rate using more species.

We find that proteins with greater dynamical influence evolve more slowly.
Moreover, we find that a protein's dynamical influence is uncorrelated with its essentiality as measured by knock-out phenotype.
We do find that in our data that evolutionary rate is also negatively correlated with essentiality, but this correlation is typically weaker than with dynamical influence.
We also examine the correlation of dynamical influence with previously established predictors of protein evolutionary rate.
In particular, we find that although dynamical influence and expression level are positively correlated, they appear to exert independent effects on evolutionary rate.

The advent of systems biology has led to much finer knowledge of the role played by specific proteins in their network contexts.
Here we have leveraged this knowledge to measure protein functional importance by the influence of that protein on the dynamics of its network.
We find that the dynamical influence of a protein is negatively correlated with its evolutionary rate and uncorrelated with its knock-out essentiality.
These observations offer a finer perspective on the role of functional importance in the protein evolution.

\section*{Methods and Materials}

\subsection*{Data and statistics}


Dynamical biochemical models encoded in the Systems Biology Markup Language~\cite{bib:Hucka2003} and formulated as systems of ordinary differential equations were obtained from the 14th release of the BioModels database~\cite{bib:LeNovere2006}.
We considered all deterministic models which involved 8 or more distinct proteins, as annotated by their UniProt~\cite{bib:Bairoch2005} identifiers, yielding 10 models.
One additional model~\cite{bib:Borisov2009} was obtained from the Molecular Systems Biology structured data archive, and another~\cite{bib:Albeck2008} was converted to SBML from published Matlab scripts.

Protein sequence alignments were obtained from the Homologene database~\cite{bib:Wheeler2007}, after mapping UniProt identifiers to UniGene identifiers. Corresponding DNA sequences were obtained from GenBank.

\begin{figure}
\centering
\includegraphics[width=2.5in]{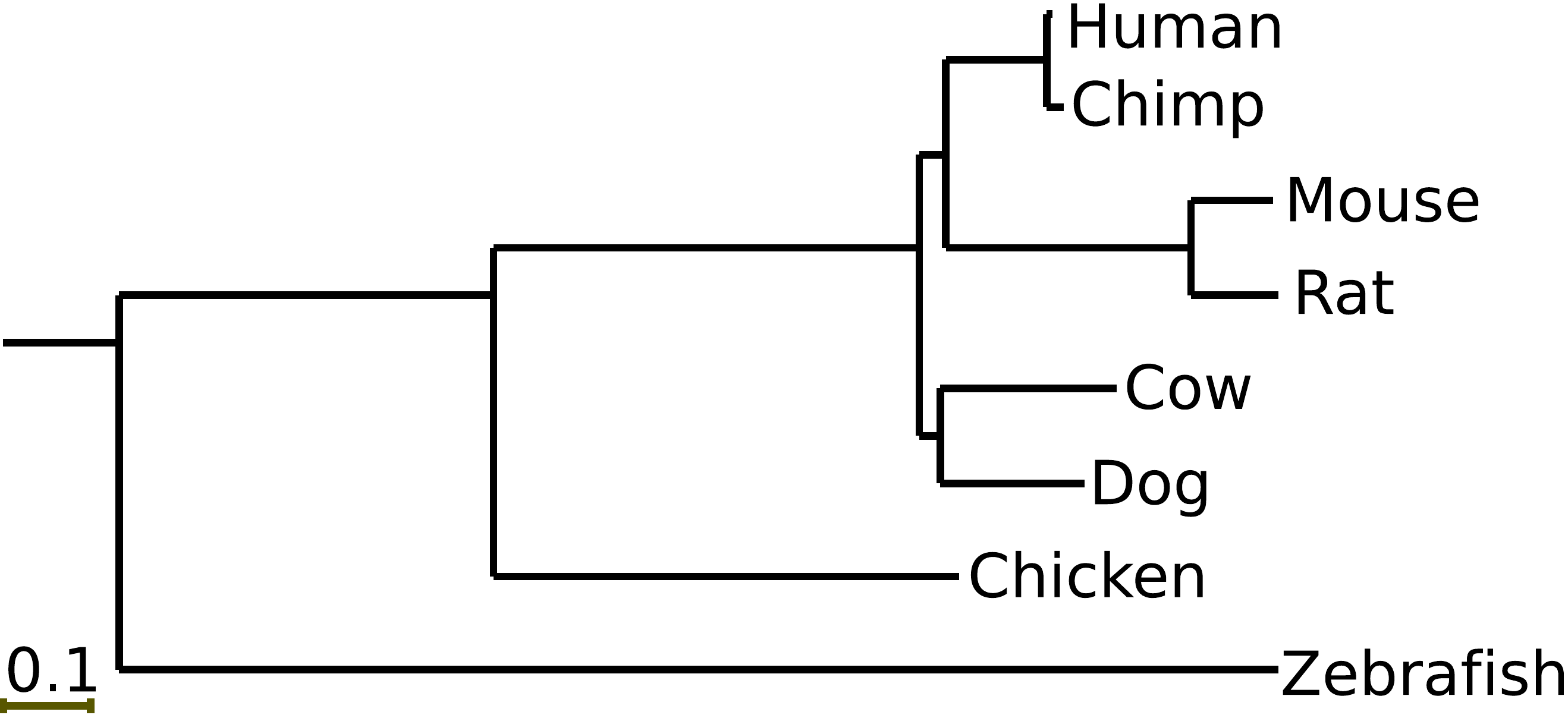}
\caption{Species tree used for measuring evolutionary rate. This tree is that obtained from concatenating all sequences studied, and the scale bar is in synonymous substitutions.\label{fig:tree}}
\end{figure}

PAML~\cite{bib:Yang2007paml} was used to obtain a maximum-likelihood estimate of dN/dS for each protein over the tree shown in Figure~\ref{fig:tree}.
Only proteins for which 3 or more sequences were available were considered.

Following~\cite{bib:Liao2006}, proteins were identified as essential or non-essential based on targeted deletion experiments compiled in the Mouse Genome Database (\url{http://www.informatics.jax.org}).
A protein was deemed essential if it resulted in a phenotype of ``lethality-prenatal/perinatal'' or ``lethality-postnatal'', including more specific classifications.
We expect this essential vs non-essential characterization to be well-conserved among vertebrates.

Mouse mRNA average expression and tissue specificity data were obtained from~\cite{bib:Liao2006}.

Correlations are assessed using Pearson's linear correlation coefficient. Reported p-values are from one-sided permutation tests.
They correspond to the probability by chance of the correlation having the expected sign and being of the observed magnitude or larger.

\subsection*{Dynamical influence}

\begin{figure}
\begin{center}
\subfigure[Example system]{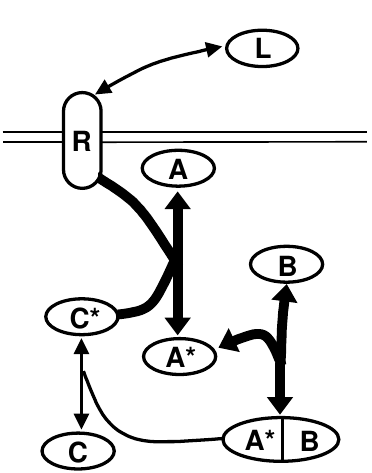}
\subfigure[Dynamical trajectories]{\includegraphics[width=1.5in]{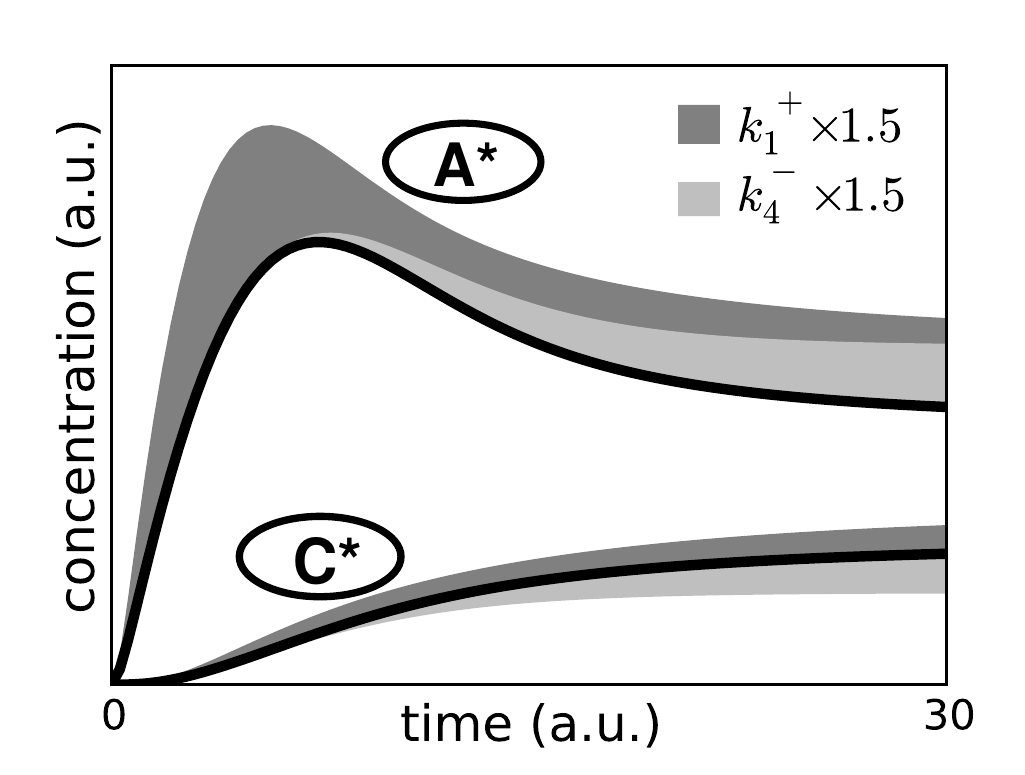}}\label{fig:trajectories}
\end{center}
\caption{(a) Reactions indicated with bold arrows in this example network are considered when calculating the dynamical influence of protein A. Note also that the external protein ligand L would not be considered in our analysis. (b) Dynamical trajectories are shown for the active forms of proteins A and C. The bold lines are trajectories given the basal rate constants, while the dark and light shaded regions indicate the change in dynamics upon increasing $k_1^+$ or $k_4^-$, respectively. In this case $k_1^+$ has a larger effect than $k_4^-$, so the dynamical influence $\kappa_1^+$ of $k_1^+$ is greater than the influence $\kappa_4^-$ of $k_4^-$.}\label{fig:illus}
\end{figure}

To quantify the change in network dynamics as the reaction rate constants $\mathbf{k}$ vary from their published values $\mathbf{k}^*$, we introduce the sensitivity measure $\kappa_i$:
\begin{equation}
\kappa_i^2 
 = \sum_{\substack{\text{species }y \\ \text{conditions } c}} \frac{1}{T_c \sigma_y^2} \int_0^{T_c} \bigg(\frac{d y_c(t, \mathbf{k})}{d \log k_i}\bigg)^2 \bigg|_{\mathbf{k} = \mathbf{k^*}} dt\label{eqn:kappa}.
\end{equation}
This measure sums over all molecular species $y$ in the model, and we sum over all conditions $c$ considered in the original model publication.
Most commonly, the conditions considered are different levels of stimulation by an external ligand.
The normalization $\sigma_y$ is taken to be the maximum of $y$ over all conditions considered. 
See Figure~\ref{fig:illus}(b) for an illustration of how varying rate constants changes dynamics.
The $\kappa_i$ were calculated using SloppyCell (\url{http://sloppycell.sourceforge.net}), and the scripts used are provided in supplementary data.

We define the  dynamical influence of a protein to be the geometric mean of the influence of all rate constants for all reactions it is involved in:
\begin{equation}
D = \left<\kappa_i\right>_\mathrm{geom}\label{eqn:I}.
\end{equation}
The $\kappa_i$ are only considered for reactions in which the protein participates by itself, not in a complex.
See Figure~\ref{fig:illus}(a) for an example.
Several models consider perturbations by externally manipulated protein ligands; these ligands
were excluded from consideration, as the manipulations considered in the models may not be physiological.
Additionally, the TNF receptor was excluded from our analysis of the apoptosis model~\cite{bib:Albeck2008}, because it was modeled non-mechanistically.

\section*{Results}

\subsection*{Proteins with greater dynamical influence evolve more slowly}

To measure the influence of a particular protein on network dynamics, we first consider the influence of each rate constant of each reaction on network dynamics.
The influence $\kappa_i$ of rate constant $i$ is defined as the summed and integrated change in \emph{all} molecular species trajectories as that rate constant is changed (Equation~\ref{eqn:kappa}).
See Figure~\ref{fig:illus}(b) for an illustration of how changes in rate constants are reflected in changes in dynamics.
We sum influence over all molecular species because it is not clear which aspects of each network's dynamics are relevant to an animal's fitness.
Results are, however, qualitatively similar if we consider changes in only a select subset of ``output'' molecular species (see online supporting data).
Given the influence of each rate constant, the influence $D$ of a protein is then defined as the geometrical mean of the influence of all rate constants for all reactions it is involved in (Figure~\ref{fig:illus}(a)).

We measure the evolutionary rate of each protein via the ratio of the rate of nonsynonymous (amino acid changing, dN) to synonymous (amino acid preserving, dS) genetic mutation, as inferred over the tree seen in Figure~\ref{fig:tree}.
(Results are qualitatively similar if we consider only the tree of mammals, rather than vertebrates (see supporting data).)

\begin{figure}
\centering
\includegraphics[width=\columnwidth]{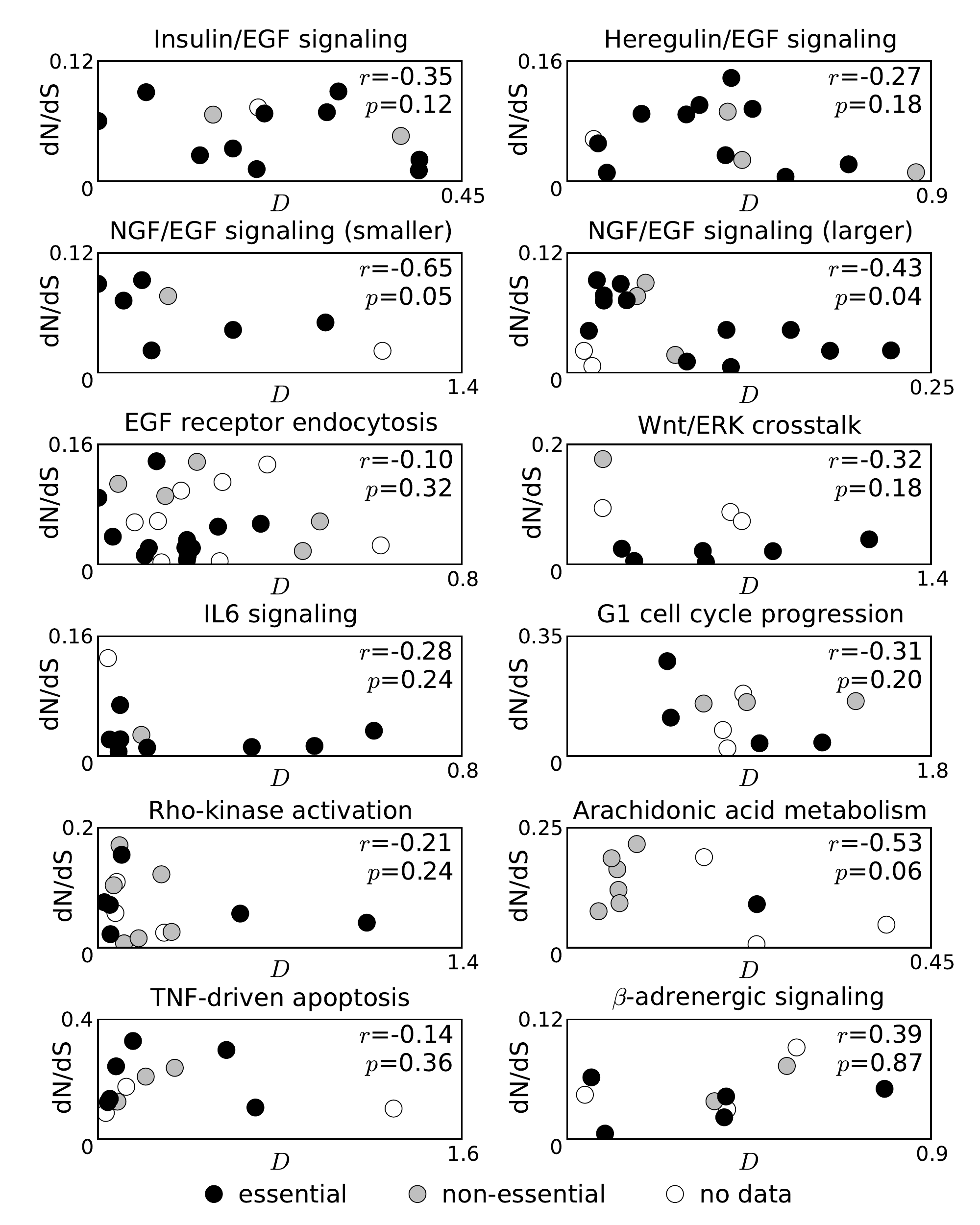}
\caption{Relationship between protein evolutionary rate and dynamical influence. Each point represents a particular protein in each system. Proteins deemed essential in mouse knock-out studies are plotted in black, non-essential in grey. There was no essentiality data available for proteins plotted in white.\label{fig:correlation}}
\end{figure}

Figure~\ref{fig:correlation} plots, for each system, protein evolutionary rate dN/dS versus protein influence $D$ on network dynamics.
We consider each system separately, because it is not clear that changes in the dynamics of, for example, arachidonic acid metabolism and ErbB signaling would yield comparable fitness effects.
Table~\ref{tbl:main} summarizes these results.

\begin{table*}
\centering
\begin{tabular}{ccccc}
system & $r_{dN/dS,D}$ (p-val, N) & $r_{dN/dS,E}$ (p-val, N) & $r_{D,E}$ (p-val, N) & $r_{dN/dS,D|E}$ \\
\hline
Insulin/EGF signaling \cite{bib:Borisov2009} & -0.35 (0.12, 13) & -0.11 (0.37, 12) & -0.14 (0.36, 12) & -0.37\\
Heregulin/EGF signaling \cite{bib:Birtwistle2007} & -0.27 (0.18, 14) & +0.20 (0.74, 13) & -0.43 (0.08, 13) & -0.24\\
NGF/EGF signaling (smaller) \cite{bib:Brown2004} & -0.65 (0.05, 8) & -0.22 (0.43, 7) & +0.05 (0.43, 7) & -0.47\\
NGF/EGF signaling (larger) \cite{bib:Sasagawa2005} & -0.43 (0.04, 17) & -0.16 (0.30, 15) & +0.17 (0.70, 15) & -0.64\\
EGF receptor endocytosis \cite{bib:Ung2008} & -0.10 (0.32, 25) & -0.44 (0.04, 17) & -0.36 (0.08, 17) & -0.44\\
Wnt/ERK crosstalk \cite{bib:Kim2007} & -0.32 (0.18, 10) & -0.98 (0.14, 7) & +0.46 (1.00, 7) & +0.65\\
IL6 signaling \cite{bib:Singh2006} & -0.28 (0.24, 10) & -0.08 (0.34, 9) & +0.18 (0.56, 9) & -0.08\\
G1 cell cycle progression \cite{bib:Haberichter2007} & -0.31 (0.20, 10) & -0.26 (0.31, 7) & -0.28 (0.25, 7) & -0.56\\
Rho-kinase activation \cite{bib:Maeda2006} & -0.21 (0.24, 16) & -0.04 (0.45, 12) & +0.26 (0.75, 12) & -0.23\\
Arachidonic acid metabolism \cite{bib:Yang2007} & -0.53 (0.06, 10) & -0.36 (0.29, 7) & +0.98 (1.00, 7) & +0.67\\
TNF-driven apoptosis \cite{bib:Albeck2008} & -0.14 (0.36, 12) & +0.09 (0.60, 9) & +0.11 (0.61, 9) & +0.08\\
$\beta$-adrenergic signaling \cite{bib:Neves2008} & +0.39 (0.87, 10) & -0.40 (0.24, 7) & -0.21 (0.34, 7) & +0.31\\
\hline
expected sign & $-$ & $-$ & $+$ & $-$ \\
 total of expected sign & 11 & 10 & 7 & 8 \\
\end{tabular}
\caption{Correlations between evolutionary rate dN/dS, dynamical influence $D$, and essentiality $E$. p-values are from one-sided permutation tests, and N is the number of proteins in each correlation.\label{tbl:main}}
\end{table*}

We expect that proteins with greater influence on network dynamics are functionally more important and thus should evolve more slowly.
For 11 of the 12 systems studied, there is indeed a negative correlation between protein evolutionary rate and dynamical influence (Table~\ref{tbl:main}).
The single exception is the $\beta$-adrenergic signaling model, which may be unreliable because it is a well-mixed approximation to a spatial model that was actually fit to data.
(Technical limitations presently preclude us from considering the partial-differential-equation spatial model in this analysis.)
Individual p-values are not dramatic; each plot includes few points because building such detailed dynamic models is currently low-throughput, and the models are thus relatively small.
If all the systems were completely independent, the odds of 11 of 12 showing a negative correlation by chance are $12/2^{12}$, which is less than 0.3\%.
The first 7 systems in Table~\ref{tbl:main} do share some proteins between them, because the EGF signaling pathway is the most often-modeled system in higher eukaryotes.
If we eliminate all proteins shared between two or more systems, then 10/11 show a negative correlation, expected by chance at the 0.5\% level.
(The smaller NGF/EGF signaling model is eliminated in this comparison because all save one of the proteins in it are shared.)
See supporting data for details.

The ratio dN/dS is affected by both changes in the substitution rate dN and in the neutral mutation rate dS.
In the supporting data, we show that the correlation between dN/dS and $D$ we observe here is not due to correlation with dS.

\subsection*{Proteins with greater dynamical influence are not more essential}

The most common measure of a protein's functional importance is essentiality, whether or not an organism can survive without it.
As in~\cite{bib:Liao2006}, we assessed essentiality by whether targeted knock-out of a protein yielded lethality in mice.
Table~\ref{tbl:main} shows that 10 of 12 systems show the expected negative correlation between essentiality $E$ and evolutionary rate, and that the correlations are typically weaker than that between dynamical influence and evolutionary rate.

Perhaps surprisingly, we find little correlation between dynamical influence and essentiality.
The correlation is only positive for only 7 of 12 models (p-value 0.39).
This suggests that the correlation we observe between $D$ and dN/dS is independent of the correlation between $E$ and dN/dS.
As seen in Table~\ref{tbl:main}, the partial correlations between evolutionary rate and dynamical influence typically remain negative when controlling for essentiality.
Of the 11 systems with a negative correlation between dN/dS and $D$, three do change sign when controlling for essentiality.
It is difficult to assess the significance of these sign flips, however, because in two of these three cases only one protein was essential or nonessential.

\subsection*{Other potential correlates}

\begin{table*}
\centering
\footnotesize
\begin{tabular}{ccccccc}
& \multicolumn{3}{c}{average expression $X$} & \multicolumn{3}{c}{expression specificity $\tau$}\\
system & $r_{dN/dS,X}$ (p-val, N)& $r_{D,X}$ (p-val, N)& $r_{dN/dS,D|X}$ & $r_{dN/dS,\tau}$ (p-val, N) & $r_{D,\tau}$ (p-val, N) & $r_{dN/dS,D|\tau}$\\
\hline
Insulin/EGF signaling \cite{bib:Borisov2009} & -0.78 (0.01, 11) & +0.77 (1.00, 11) & +0.42 & +0.21 (0.71, 11) & -0.08 (0.40, 11) & -0.43\\
Heregulin/EGF signaling \cite{bib:Birtwistle2007} & -0.45 (0.08, 10) & +0.38 (0.85, 10) & -0.07 & +0.14 (0.66, 10) & -0.05 (0.47, 10) & -0.23\\
NGF/EGF signaling (smaller) \cite{bib:Brown2004} & -0.39 (0.22, 5) & +0.69 (0.88, 5) & -0.88 & +0.84 (0.98, 5) & -0.66 (0.11, 5) & -0.73\\
NGF/EGF signaling (larger) \cite{bib:Sasagawa2005} & -0.53 (0.03, 12) & +0.50 (0.94, 12) & -0.46 & +0.44 (0.92, 12) & -0.28 (0.19, 12) & -0.56\\
EGF receptor endocytosis \cite{bib:Ung2008} & -0.43 (0.01, 23) & +0.19 (0.82, 23) & -0.14 & +0.43 (0.98, 23) & -0.00 (0.50, 23) & -0.23\\
Wnt/ERK crosstalk \cite{bib:Kim2007} & -0.87 (0.00, 7) & -0.38 (0.19, 7) & +0.00 & +0.17 (0.67, 7) & +0.83 (0.98, 7) & +0.34\\
IL6 signaling \cite{bib:Singh2006} & -0.58 (0.02, 9) & -0.12 (0.39, 9) & -0.47 & +0.16 (0.72, 9) & -0.10 (0.45, 9) & -0.30\\
G1 cell cycle progression \cite{bib:Haberichter2007} & -0.07 (0.43, 10) & +0.46 (0.90, 10) & -0.31 & +0.55 (0.95, 10) & -0.43 (0.11, 10) & -0.10\\
Rho-kinase activation \cite{bib:Maeda2006} & -0.09 (0.39, 13) & +0.09 (0.66, 13) & -0.15 & +0.45 (0.93, 13) & +0.01 (0.56, 13) & -0.18\\
Arachidonic acid metabolism \cite{bib:Yang2007} & -0.54 (0.12, 7) & +0.47 (0.77, 7) & -0.23 & +0.00 (0.51, 7) & -0.06 (0.58, 7) & -0.43\\
TNF-driven apoptosis \cite{bib:Albeck2008} & -0.50 (0.08, 9) & -0.04 (0.43, 9) & -0.39 & +0.49 (0.90, 9) & -0.28 (0.22, 9) & -0.22\\
$\beta$-adrenergic signaling \cite{bib:Neves2008} & -0.66 (0.03, 8) & -0.44 (0.13, 8) & +0.42 & +0.09 (0.59, 8) & +0.17 (0.66, 8) & +0.57\\
\hline
expected sign & $-$ & $+$ & $-$ & $+$ & $-$ & $-$ \\
 total of expected sign & 12 & 8 & 9 & 12 & 9 & 10 \\
\end{tabular}
\caption{Correlations between evolutionary rate dN/dS, dynamical influence D, and expression-related variables. p-values are from one-sided permutation tests, and N is the number of proteins in each correlation\label{tbl:other}}
\end{table*}

The most well-established correlate with protein evolutionary rate is level of expression. Table~\ref{tbl:other} reports correlations in our data between evolutionary rate and expression-related variables.

To measure expression level, we used the average expression level across 61 mouse tissues, as compiled by~\cite{bib:Liao2006}.
As found in other studies~\cite{bib:Drummond2005}, evolutionary rate is consistently negatively correlated with expression level in our data (Table~\ref{tbl:other}).
Moreover, dynamical influence and expression level are positively correlated in 8 of 12 cases.
Of the 11 systems for which evolutionary rate and dynamical influence are negatively correlated, in only 2 cases does controlling for expression level via partial correlation change the sign.
This suggests that dynamical influence and expression level, although correlated with each other, exert mostly independent effects on evolutionary rate.

In vertebrates, expression level may vary from tissue to tissue, so we also we consider data on tissue specificity $\tau$ compiled by~\cite{bib:Liao2006}.
Tissue specificity is consistently positively correlated with evolutionary rate (Table~\ref{tbl:other}); proteins expressed in few tissues evolve faster.
Dynamical influence and specificity are negatively correlated in 9 of 12 cases, indicating that influential proteins tend to be expressed in many tissues.
Controlling for tissue specificity, dynamical influence and evolutionary rate remain negatively correlated for 10 of the 11 systems for which they began negatively correlated, suggesting that these factors independently influence the rate of protein evolution.

We have also examined correlations with gene compactness~\cite{bib:Liao2006} and the number of reactions a protein participates in (see supporting data).
We find no consistent correlation between a protein's dynamical influence and it's genetic compactness.
We do find that there is a negative correlation between a protein's dynamical influence and the number of reactions it participates in.
In both cases, however, it appears that dynamical influence exerts an independent effect on evolutionary rate.

\section*{Discussion}

We have introduced a fine scale measure of protein functional importance, based on the influence each protein has on the dynamics of the biochemical network in which it is embedded.
We have shown that in the systems we study proteins with greater dynamical influence evolve more slowly, although they are no more essential as determined by mouse knock-outs.
Moreover, the correlation of dynamical influence with evolutionary rate appears to be independent of expression level and specificity.
It also appears to be independent of gene compactness and number of reactions a protein is involved in.

Of the 12 systems tested, 11 show a negative correlation between protein dynamical influence and evolutionary rate.
The one exception, a model of $\beta$-adrenergic signaling, is a well-mixed approximation to a spatial model for which we do not currently have the techniques to analyze parameter sensitivity, and it was the spatial model that was actually fit to biochemical data.
The approximate model may thus not properly reflect the actual network dynamics, explaining why this system is an exception.
This emphasizes the importance of model quality in assessing dynamical influence.
We measured dynamical influence using computational models because it is at present technically very challenging to experimentally introduce small controlled perturbations to protein activity.
All models, however, are approximations to reality, which introduces noise into our measure of dynamical influence.
In particular, the models currently available of necessity omit many potential network components and interactions.
These omissions may explain the many cases in which a protein has low measured dynamical influence but evolves slowly. 
(Points clustered near 0,0 in Figure~\ref{fig:correlation}).
The rate constant values used in the current models are also likely inaccurate.
It appears, however, that these models nevertheless capture the correct sensitivity to changes in rate constants.
As more complete and refined models are developed, we may find that the observed correlation between dynamical influence and evolutionary rate grows stronger.

Many proteins are modular, with different domains performing different reactions.
Here we measure sequence divergence of whole proteins and define dynamical influence using all reactions in which a protein is involved.
Considering the evolution of each protein domain separately and computing its influence using only the reactions it is actively involved in may make our analysis more powerful.
Similarly, we eliminated protein complexes from analysis, but future studies may identify which components of a complex participate in which reactions.
Both these refinements are difficult at present, because models are not typically annotated with the necessary information.
As domain-oriented rules-based models~\cite{bib:Hlavacek2006} grow in prominence, this limitation may disappear.

We measure protein dynamical influence by first assessing how changing each reaction rate constant affects network dynamics, and we measure evolutionary rate by sequence divergence.
However, changes in a protein sequence do not necessarily correspond to changes in rate constants~\cite{bib:Kiel2008}.
Ideally, we would directly measure the evolutionary divergence in rate constants, but at present these quantities are difficult to measure in even a single organism.
Computational methods that predict the impact of particular sequence changes on protein structure (e.g.~\cite{bib:Sunyaev2001}) may provide a more relevant measure of evolutionary change.
However, given that that we see a correlation between dynamical influence and protein sequence evolution, we expect that the correlation between dynamical influence and more direct measures of rate constant evolution would be even stronger.

A further underlying assumption in our analysis is that the structure of the network remains constant over the evolutionary distances considered.
A recent study of network structure evolution in eukaryotes indeed found that it is typically well-conserved among vertebrates.
The stability of the networks we consider is further supported by the fact that the correlations we observe are stronger when the tree of vertebrates is used rather than the tree of mammals.
The reduction in noise resulting from using more sequence in assessing dN/dS thus appears to outweigh noise from possible changes in network structure or function.

Here we have examined evolutionary rate using biochemical pathway models.
Other authors have also taken a pathway perspective evolutionary rate.
Ramsay et al.\ showed that proteins acting earlier in particular plant metabolic synthesis pathways evolved more slowly than those downstream~\cite{bib:Ramsay2009}.
Similarly, Greenberg et al.\ showed that proteins in the metabolic ``core'' or that share metabolites with other enzymes evolve more slowly~\cite{bib:Greenberg2008}.
Other authors have used flux-balance analysis to consider the steady-state properties of the yeast metabolic network:
Vitkup et al.\ showed that proteins mediating greater metabolic flux evolved more slowly~\cite{bib:Vitkup2006}, while Wang et al. showed that proteins with larger knock-out effect on the steady-steady metabolic network evolve more slowly~\cite{bib:Wang2009}.
Here we consider not only structure or steady-state behavior, but also dynamics.
We also consider a different class of networks, as most of our examples are signaling networks, the exceptions being models of the cell cycle and arachidonic acid metabolism.

It is perhaps surprising that in these systems dynamical influence is independent of knock-out essentiality.
This may reflect the presence of many reactions which must take place for survival, but whose precise rate matters little, perhaps due to other forms of regulation.
Proteins involved in such reactions would have dramatic effects if knocked out (and thus be deemed essential), but the network would be relatively insensitive to changes in their sequence (and thus they would have low dynamical influence).

A great deal of effort has been devoted to teasing apart which correlates of evolutionary rate exert independent effects.
This has proven very difficult, and there is no well-established statistical test that may be used.
Drummond et al.\ have shown that partial correlation may overestimate independence between noisy factors, and they argue in favor of principle components regression~\cite{bib:Drummond2006}.
Principle components regression, however, has similar difficulties when variables differ in their noisiness~\cite{bib:Plotkin2007}.
Here we have used partial correlation analysis, as it is simple and can be applied to the small number of data points we have for each system.
Our results are thus suggestive of independent effects between dynamical influence and other factors, but they are not definitive.
It does seem, however, that essentiality and dynamical influence are themselves almost uncorrelated, implying that their effects on evolutionary rate are independent.

We have introduced a measure of protein functional importance, dynamical influence, that assesses the effects of small changes in protein reaction efficacy on network dynamics.
Dynamical influence is negatively correlated with evolutionary rate as measured by dN/dS, but it is independent of essentiality, a commonly used but much coarser measure of importance.
By taking a finer view of protein importance, our results shed light on the controversy as to how much functional importance matters for evolution, suggesting that more important proteins do indeed evolve more slowly.

\begin{acknowledgments}
I thank Sarah Stockwell, Tony Greenberg, Tanja Kortemme and particularly Simon Gravel for helpful comments and suggestions. I also thank Jianzhi Zhang and Ben-Yang Liao for graciously sharing their compiled expression data.
\end{acknowledgments}




\bibliography{../ProteinEvol}

\end{document}